\toks1={Time-stamp: <27-Nov-2015 01:06:02 adk@BIGBENJAMIN>}
\documentclass{PoS}
\title{Renormalizability of the Schrödinger Functional}
\ShortTitle{Renormalizability of the Schrödinger Functional}
\author{\speaker{A. D. Kennedy}\\
  School of Physics \& Astronomy, University of Edinburgh, Scotland\\
  E-mail: \email{adk@ph.ed.ac.uk}}
\author{{Stefan Sint}\\
  School of Mathematics, Trinity College Dublin, Dublin 2, Ireland\\
  E-mail: \email{sint@maths.tcd.ie}}
\abstract{Symanzik showed that quantum field theory can be formulated on a
  space with boundaries by including suitable surface interactions in the
  action to implement boundary conditions. We show that to all orders in
  perturbation theory all the divergences induced by these surface
  interactions can be absorbed by a renormalization of their coefficients.}
\FullConference{The 33rd International Symposium on Lattice Field Theory\\
  14--18 July 2015\\
  Kobe International Conference Center, Kobe, Japan}
\usepackage{epsfig}
\def\epspdffile#1{\leavevmode\ifpdf\epsffile{#1.pdf}\else\epsffile{#1.eps}\fi}
\def\sgn{\mathop{\rm sgn}}            
\def\implies{\Rightarrow}	      
\def\L{{\cal L}}                      
\def\rational#1#2{{\mathchoice{\textstyle{#1\over#2}}%
  {\scriptstyle{#1\over#2}}{\scriptscriptstyle{#1\over#2}}{#1/#2}}}
\def\half{\rational12}		      
\def\formattoks Time-stamp: <#1>{#1}
\def\ip(#1,#2){#1{\cdot}#2}           
\begin{document}

\section{Introduction}

It is of interest to study quantum field theories with boundaries for a
variety of reasons: the Casimir effect~\cite{Casimir} describes a quantized
electromagnetic field with mirrors as spatial boundaries; and the
Schr\"odinger functional~\cite{Symanzik} describes a quantum field theory with
specified initial and final field configurations, which has interesting
applications to renormalization problems in lattice gauge theories and lattice
QCD~\cite{Luscher, Sint}.  Symanzik showed that such boundary conditions may
be imposed by including suitable surface interactions in the
action~\cite{Symanzik}.  Our goal is to establish that quantum field theories
with boundary conditions imposed in this way are renormalizable to all orders
in perturbation theory.

Our proof is independent of the choice of regulator: we shall use continuum
notation for simplicity, but results apply equally well with a lattice
regulator.  The proof is in Euclidean space; extension to Minkowski space,
where the Green's functions are distributions rather than functions presumably
follows by partial integration onto test functions just as for theories
without boundaries~\cite{Lowenstein}.  We shall only present a proof for a
scalar field here.

\section{Boundary Conditions}

For simplicity in this summary we describe the proof for a scalar field
\(\phi\) with the Lagrangian density \(\L = \half(\partial\phi)^2 + \half
m^2\phi^2 + \rational1{4!}\lambda\phi^4\) to which we add the surface term \(K
= \half c\phi_{-}\delta'(\sigma)\phi_{+}\), with \(c = \pm1\) so as to
decouple the two sides of the wall.  This is similar but not identical to
Symanzik's surface interaction.  The function \(\sigma\) vanishes on the wall,
and specifies the location of the boundary.  For a planar wall that is
orthogonal to a unit vector \(w\) and a distance~\(\ell\) from the origin we
could take \(\sigma(x) = \ip(x,w) - \ell\).  In general we can take \(\sigma\)
to be a smooth function corresponding to a wall that is topologically
equivalent to a plane.

\subsection{Quadratic Interactions}

Observe that the boundary conditions are imposed by a local interaction that
is quadratic in the field~\(\phi\), and moreover there is no small parameter
associated with this wall interaction.  This is analogous to the mass term
\(\half m^2\phi^2\): we can either treat this as part of the propagator,
\((k^2 + m^2)^{-1}\), or treat it ``perturbatively'' as a two-point vertex
\(-m^2\) with the massless propagator \(\Delta = 1/k^2\).  In the latter case
we can sum the two-point interactions to all orders in~\(m\)
\begin{eqnarray}
  \Delta_M &=& \Delta + \Delta (-m^2) \Delta
    + \Delta (-m^2) \Delta (-m^2) \Delta + \cdots 
  = \Delta \sum_{n=0}^\infty \left[(-m^2) \Delta\right]^n \nonumber \\
  &=& \Delta + \Delta (-m^2) \Delta_M
  = \frac\Delta{1 + m^2\Delta}
  = \frac1{k^2 + m^2}.
\end{eqnarray}
Of course, this series only converges for \(k^2 \leq m^2\), but it has a
unique analytic continuation \(\forall k^2 \neq -m^2\), even though there is
no small parameter.  

This equivalence should be familiar: the mass renormalization is \(m^2 \to m^2
+ \delta m^2\), where \(\delta m^2\) is treated as a countervertex order by
order in the loop expansion.

\subsection{Integral Equation}

The Green's function \(H(x,y)\) for the quadratic kernel without walls
\begin{equation}
  L(x,y) = \delta(x - y)(-\partial^2 + m^2)
\end{equation}
satisfies \(\int dz\, L(x,z) H(z,y) = \delta(x, y)\), which we shall
abbreviate as \(LH = 1\).  Of course, we must also specify suitable boundary
conditions to uniquely specify the Green's function, so We require that
\(\lim_{|x - y|\to\infty} H(x - y) = 0\).  Because there are no walls \(H\) is
translationally invariant and is only a function of \(x - y\).  The Green's
function \(G(x,y)\) for the full quadratic kernel \(L+K\) where the wall
interaction is
\begin{equation}
  K(x,y) = \int dz\, \delta(x - z_{-}) \delta(y - z_{+}) \delta'(\sigma(z))
\end{equation}
which satisfies \((L + K) G = 1\), where \(z_{\pm} = z \pm \varepsilon
\partial\sigma\) with \(\varepsilon\to0\).  Naturally we also chose the
boundary conditions that \(G(x,y) = 0\) as \(|x|\to\infty\) or
\(|y|\to\infty\).

Following Symanzik we may thus find \(G\) ``non-perturbatively'' by solving
the integral equation \((L+K)G = 1\); upon multiplying on the left by \(H\)
this gives \(H (L + K) G = H \implies (1 + HK) G = H\).  \(G(x,y)\) is not
translationally invariant, so it is not just a function of~\(x - y\).  We
require \(G(x_{-},x_{+}) = 0\), so the two sides of the wall are decoupled;
moreover all the propagator's derivatives must also vanish across the wall.
Since the propagator \(G\) vanishes across the wall so does any connected
Green's function that couples points on opposite sides of the wall, as it is a
convolution of propagators.

For simplicity we now assume the wall is the plane orthogonal to the \(x_1\)
axis and passing through the origin.  We may then reduce the problem of
finding the Green's function \(G\) to a one-dimensional one by considering a
single Fourier mode in the \(D-1\) directions parallel to the wall with
frequency \(\omega\).

Inserting the explicit solution \(H(x_1,y_1) = e^{-\omega|x_1-y_1|}/2\omega\)
this leads to the linear system
\begin{equation}
  \left( \begin{array}{cccc}
    1 - \frac c2 & -\frac c2 & -\frac c{2\omega} & -\frac c{2\omega} \\
    \frac c2 & 1 + \frac c2 & -\frac c{2\omega} & -\frac c{2\omega} \\
    \frac{c\omega}2 & \frac{c\omega}2 & 1 + \frac c2 & \frac c2 \\
    \frac{c\omega}2 & \frac{c\omega}2 & -\frac c2 & 1 - \frac c2
  \end{array} \right) \left( \begin{array}{c}
    G(x_1,0_+) \\ G(x_1,0_-) \\ \partial_2G(x_1, 0_+) \\ \partial_2G(x_1,0_-)
  \end{array} \right) = \frac{e^{-\omega|x_1|}}{2\omega} \left( \begin{array}{c}
    1 \\ 1 \\ \omega\sgn x_1 \\ \omega\sgn x_1
  \end{array} \right),
\end{equation}
where \(\partial_2G\) stands for the derivative of \(G\) with respect to its
second argument, and we have used the fact that \(G\) and its derivatives are
left (right) continuous on the left (right) of the wall.

\subsection{Solution of Integral Equation}

\begin{figure}[ht]
  \begin{center}
    {\epsfxsize=0.7\textwidth \epspdffile{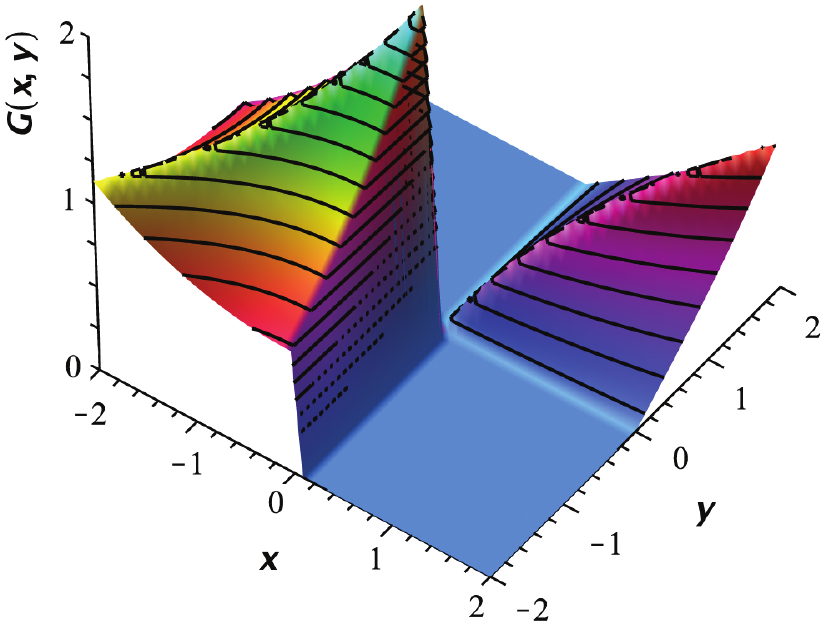}}
  \end{center}
  \caption{The Green's function \(G(x_1,y_1)\) with \(\omega=\half\) and
    \(c=-1\) for a scalar field with a wall interation \(K\) at \(\sigma(x) =
    x_1 = 0\).  This Green's function is not translationally invariant, so it
    is not just a function of \(x_1-y_1\).  It satisfies Dirichlet boundary
    conditions on the right and Neumann boundary conditions on the left.}
  \label{fig:1}
\end{figure}

We find that the two sides of the wall only decouple for \(c=\pm1\), which
crucially does not depend on the frequency~\(\omega\).  The solution of the
integral equation for \(c=-1\) is
\begin{equation}
  \omega G(x_1,y_1) = \left\{ \begin{array}{l@{\qquad}l}
    e^{\omega y_1} \cosh\omega x_1 & \mbox{for \(y_1 \leq x_1 \leq 0\)} \\
    e^{\omega x_1} \cosh\omega y_1 & \mbox{for \(x_1 < y_1 \leq 0\)} \\
    0 & \mbox{for \(x_1 \leq 0 < y_1\)} \\
    0 & \mbox{for \(y_1 \leq 0 < x_1\)} \\
    e^{-\omega x_1} \sinh\omega y_1 & \mbox{for \(0 < y_1 \leq x_1\)} \\
    e^{-\omega y_1} \sinh\omega x_1 & \mbox{for \(0 < x_1 < y_1\).}
  \end{array} \right.    
\end{equation}

This solution for \(\omega = \half\) is shown in Figure~\ref{fig:1}.  It
satisifies Dirichlet boundary conditions on the right and Neumann boundary
conditions on the left.  Changing the sign of the wall interaction to \(c=1\)
interchanges these.

\section{Divergences and Renormalization}

\subsection{Feynman Rules}

\begin{figure}[ht]
  \begin{center}
    {\epsfxsize=0.35\textwidth \epspdffile{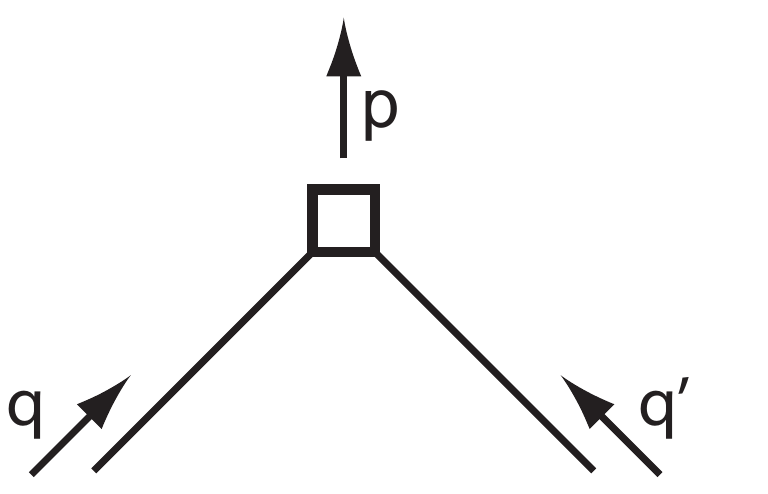}}
  \end{center}
  \caption{Two-point vertex coupling the field to the external source, which
    is localized on the wall.}
  \label{fig:2}
\end{figure}

As well as the usual bulk divergences we have new divergences associated with
wall vertices.  For simplicity we consider the wall \(\sigma(x) = x_1 - \ell\)
which is orthogonal to the \(x_1\) axis and intersects it at \(x_1 = \ell\).
The wall vertex is thus
\begin{equation}
  K(x,y) = \int dz\, \delta(x - z_{-}) \delta(y - z_{+}) \delta'(z_1 - \ell);
\end{equation}
in momentum space this is
\begin{eqnarray}
  \tilde K(q,q') &=& \int \frac{dx\,dy}{(2\pi)^D}\,
    K(x,y) e^{-i(\ip(q,x) + \ip(q',y))}
  = \int \frac{dz}{(2\pi)^D}\, e^{-i(\ip(q,z_{-}) + \ip(q',z_{+}))} 
    \delta'(z_1 - \ell) \nonumber \\
  &=& \frac i{2\pi} (q + q')_1 e^{-i\ell(q + q')_1}
  e^{i\varepsilon(q-q')_1} \delta((q + q')_\perp) \nonumber\\
  &=& \frac i{2\pi} \int dp\, p_1 e^{-i\ell p_1} \delta(q + q' - p)
  \delta(p_\perp) e^{i\varepsilon'\sgn(q-q')_1}.
\end{eqnarray}
The location of the wall is specified by the phase \(e^{-i\ell p_1}\), and its
orientation by the dependence on the sign of \((q-q')_1\).  We have associated
an ``external'' momentum \(p\) with the wall source so that momentum is
conserved at the wall vertex; this corresponds to the two-point vertex
coupling the field the external source shown in Figure~\ref{fig:2}.

\subsection{Single Wall Vertex}

\begin{figure}
  \begin{center}
    {\epsfxsize=0.4\textwidth \epspdffile{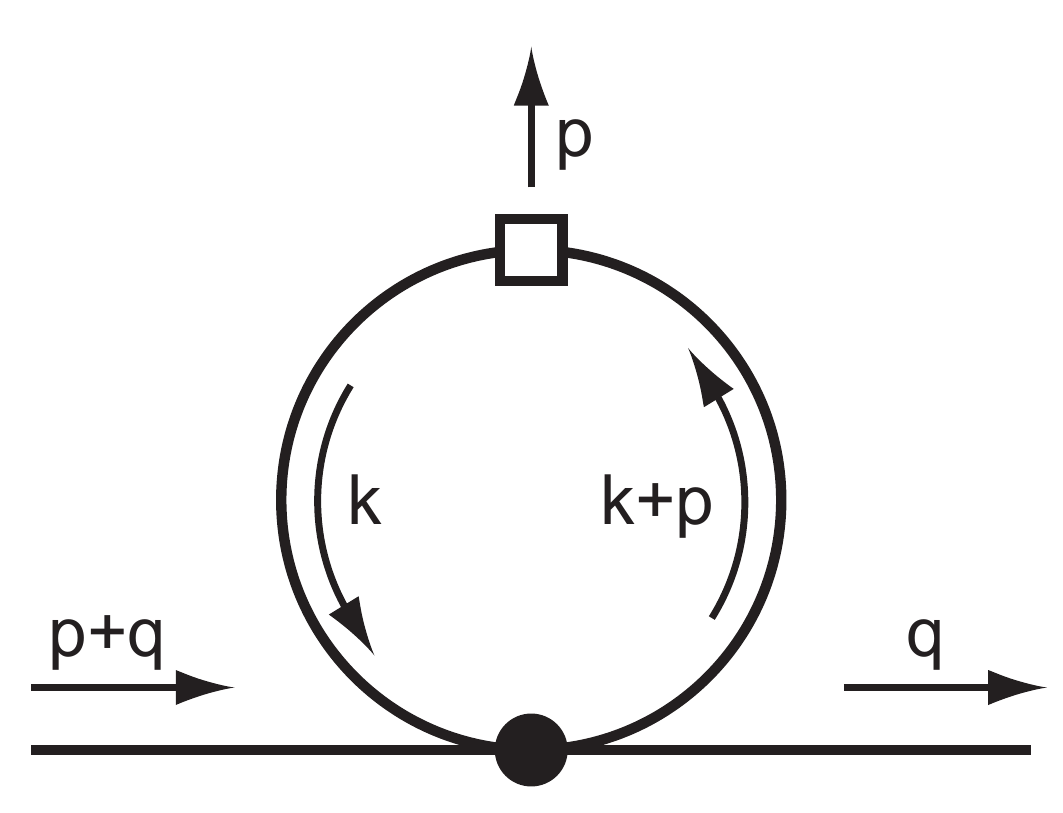}}
  \end{center}
  \caption{One-loop contribution to the two-point function that includes a
    single wall vertex.}
  \label{fig:3}
\end{figure}

Consider the one-loop diagram of Figure~\ref{fig:3} contributing to the
two-point function that includes a single wall vertex.  This is
logarithmically divergent in \(D=4\) dimensions, and therefore its divergent
part is independent of \(q\) and is proportional to the wall vertex \(\tilde
K(p+q,q)\).  This divergence may be absorbed into a renormalization of the
coefficient of the wall vertex, \(c\to c + \delta c\).  We may impose the
renormalization condition that the finite part of \(\delta c\) vanishes so as
to maintain the decoupling of the two sides of the wall.

\subsection{Multiple Wall Vertices}

\begin{figure}
  \begin{center}
    {\epsfxsize=0.4\textwidth \epspdffile{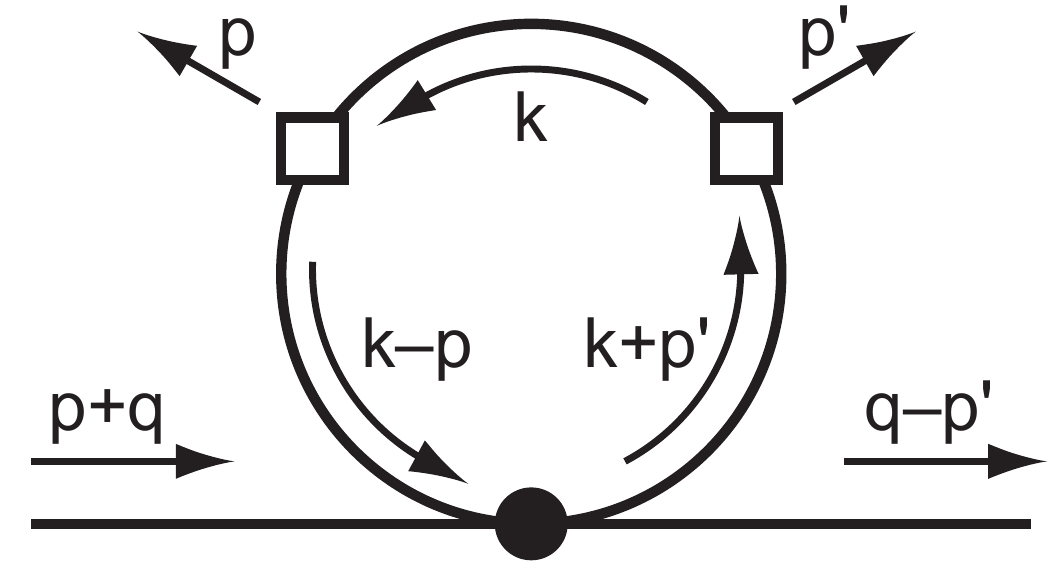}}
  \end{center}
  \caption{One-loop contribution to the two-point function that includes two
    wall vertices.}
  \label{fig:4}
\end{figure}

Consider the one-loop diagram of Figure~\ref{fig:4} contributing to the
two-point function that includes two wall vertices.  This is logarithmically
divergent in \(D=6\) dimensions (do not be distracted by the fact that
\(\phi^4\) theory without walls is not renormalizable in six dimensions), and
therefore its divergent part is independent of \(q\), but it is \emph{not\/}
proportional to a single wall vertex.  Therefore this divergence is
\emph{not\/} localized on the wall, and cannot be absorbed into a
renormalization of the coefficient of the wall vertex.

In general, if more than one wall vertex appears in an overall divergent graph
then the divergence is not localized on the wall.

\subsection{Power Counting}

We can easily apply Dyson's power-counting threorem to wall vertices.  In our
example the wall interaction monomial in the action in \(D\)~dimensions has
dimension \(D-2\), just like a mass term.  Therefore for \(D=4\) the only
overall divergent \(n\)-point functions with one wall vertex must have
\(n\leq2\).  If there are two or more wall vertices then \(n\leq0\).  \(n=1\)
is forbidden by \(\phi\to-\phi\) symmetry, and \(n=0\) is uninteresting, so
the only new counterterm required is proportional to the wall vertex and is
therefore localized on the wall.

\begin{figure}
  \begin{center}
    \dimen0=0.3\textwidth%
    {\epsfysize=1.35\dimen0 \epspdffile{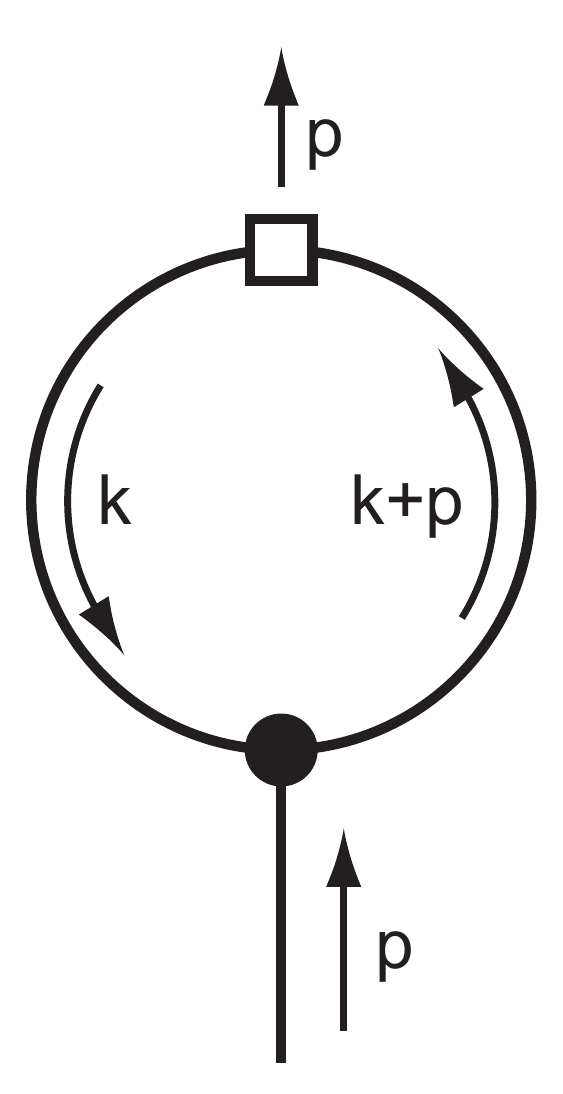}}%
    \hskip4em%
    {\epsfysize=1.1\dimen0 \epspdffile{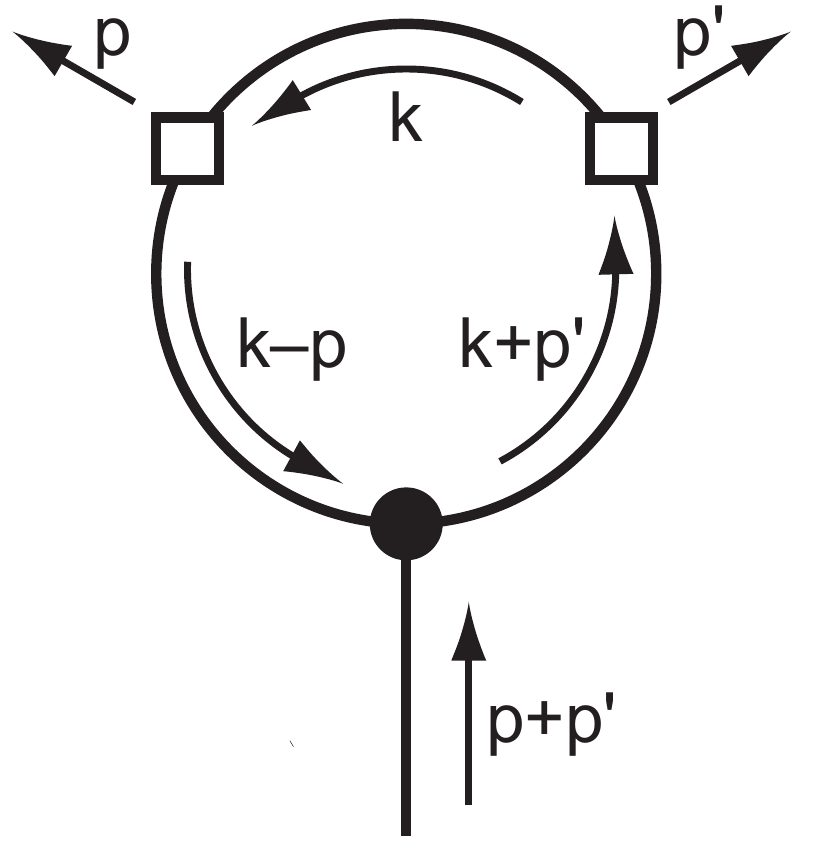}}
  \end{center}
  \caption{One-loop ``tadpole'' contributions to the one-point function.}
  \label{fig:5}
\end{figure}

The analysis for multi-loop diagrams is essentially the same.  After removing
all subdivergences using Bogoliubov's \(\bar R\)-operation there are only
logarithmic divergences involving a single wall vertex for \(D=4\), which just
gives a higher-order contribution to the counterterm \(\delta c\) in the loop
expansion.

Observe that if we were to consider \(\phi^3\) theory instead of \(\phi^4\)
theory then we would also need to consider divergent tadpoles such as the
one-loop diagrams of Figure~\ref{fig:5}. These contribute to a non-uniform
background source \(J(x)\) for the field~\(\phi\), but do not lead to any
coupling of the opposite sides of the wall.

\section{Conclusions}

Momentum is not conserved at a wall vertex: this is not suprising, as the wall
violates translational invariance. This corresponds to an ``external''
momentum \(p\) flowing into the wall; there is an integral of all possible
values of \(p\) with a uniform distribution corresponding to the Fourier
transform of the \(\delta'\) function ``shape'' of the wall. 

The wall interaction monomial in the action always has the same power-counting
dimension as the mass term.  Imposition of boundary conditions on the field by
a local wall interaction induces counterterms that remain localized on the
wall to all orders in perturbation theory provided that no more than one wall
vertex can appear in any overall divergent two-point function.

\section*{Acknowledgements}

ADK is funded by an STFC Consolidated Grant ST/J000329/1. S.~Sint acknowledges
support by SFI under grant 11/RFP/PHY3218.

\end{document}